\documentclass[12pt,a4paper]{article}
\newtheorem{defn}{Definition}[section]

\usepackage{amsmath}
\usepackage{amssymb}
\usepackage{array}
\usepackage{graphicx}
\usepackage{geometry}
\usepackage{graphics}
\usepackage{enumerate}
\usepackage{lscape}
\usepackage[utf8]{inputenc}
\usepackage{float}
\usepackage{caption} 
\author{Gladwin James V.$^1$, Thomas Xavier$^{2*}$ and Nicy Sebastian$^1$    \\
	\small{$^1$Department of Statistics,
		St.Thomas College (Autonomous), Thrissur}\\
		\small{ University of  Calicut, Kerala - 680 001, India.}\\
		\small {gladwinvjames@gmail.com, nicycms@gmail.com}\\
			\small{$^2$Theoretical Statistics and Mathematics Unit}\\
			 \small{Indian Statistical Institute, New Delhi - 110 016, India.}	\\
	\small {thomasmxavier@gmail.com}\\ \small {$^*$Corresponding author}
}
\date{}
\begin{document}
	\title{Stress-strength reliability estimation for type-1 pathway generated exponential distribution with applications to AIDS incubation time}
\maketitle
\begin{abstract}
The Type-1 Pathway Generated Exponential distribution (PGE-1) is introduced. We have considered the estimate of the stress-strength parameter when the stress and strength components are statistically independent and follow PGE-1 distributions with distinct parameters. The point estimate of the stress-strength reliability is obtained using maximum likelihood method. The interval estimates are obtained using asymptotic confidence intervals and the parametric bootstrap method. To verify the performance of the methods developed, an extensive Monte Carlo simulation study has been conducted. The application of the developed results is illustrated on AIDS blood transfusion data.

\end{abstract}
\section{Introduction}

Let $X$ and $Y$ be two statistically independent continuous random variables, then a lot of interest has been given to study the statistical inference of $R= Pr\{X > Y\}$ or which is also known as the stress-strength parameter. In engineering, $R$ measures the reliability of a system where $X$ denote the random strength and $Y$ denotes the random stress applied on the system. In medicine, $R$ can be interpreted as measure of a treatments effectiveness over the control, if $X$ and $Y$ represents the response variables from treatment and control groups respectively.

Birnbaum (1956) presented the core concept, and Birnbaum and McCarty developed it (1958). A thorough account of the evolution of stress-strength models up to that point is given by Kotz $et ~ al.$ (2003), which covers all the major distributions including exponential, normal, gamma, Weibull, Burr, generalized exponential, generalized Weibull, generalized logistic, and many more. Some comprehensive studies in stress-strength can be found in Church and Harris (1970), Kundu (2005) and Kundu and Raqab (2009). Some of the applications of the stress-strength model are discussed in Pakdaman $et ~ al.$ (2017). For the most recent results on the topic, see Xavier and Jose (2021), Pak $et ~ al.$ (2021), and Chacko and Mathew (2021). 

The pathway model was first proposed by Mathai (2005) for the rectangular matrix-variate scenario. The Pathway model lets us to switch from one functional model to other. The generalized type-1 beta model, generalized type-2 beta model, and generalized gamma model are all connected by the model. This feature is helpful when the distribution used by the model must have a thicker or thinner tail than those provided by the parametric family. Mathai and Provost (2006) extended the Pathway model to the complex domain.
A special case of the pathway model for real scalar positive variables  has the following structure:
\begin{equation}\label{p2.1}
f_{1}(x)=a \delta(\eta+1-q) x^{\delta-1} \left[1-a(1-q) x^{\delta} \right]^{\frac{\eta}{1-q}},
\end{equation}
for  $q<1, a>0, \delta>0, \eta>0, \eta+1-q>0, 1-a(1-q) x^{\delta}>0$ and $f_1(x)$ = 0 elsewhere.

\begin{equation}\label{p2.2}
f_{2}(x)=a \delta(\eta+1-q) x^{\delta-1}\left[1+a(q-1) x^{\delta}\right]^{-\frac{\eta}{q-1}},
\end{equation}
for $q>1, a>0, \delta>0, \eta>0, \eta+1-q>0, x \geq 0$ and $f_2(x)$ = 0 elsewhere. When $q\rightarrow 1_{-}$ in \eqref{p2.1} or $q\rightarrow 1_{+}$ in \eqref{p2.2}, then

\begin{equation}\label{p2.3}
f_{3}(x)=a \delta \eta x^{\delta-1} \mathrm{e}^{-a \eta x^{\delta}},
\end{equation}
for $a>0, \delta>0, \eta>0, x \geq 0$ and $f_3(x)$ = 0 elsewhere.

Mathai and Princy (2016) developed the pathway generated family and studied its various properties. For the sake of completion, the definition of the pathway-generated family of Mathai and Princy (2016) is given below: 

	\begin{defn}

Let $Y$ be a continuous random variable with distribution function $G(y)$ and density function $g(y)$ then

\begin{equation}\label{p.1}
g_{1}(y)=(\eta + 1 - q)a\delta g(y){G(y)}^{\delta-1} \left[1-a(1-q){G(y)}^{\delta} \right]^{\frac{\eta}{1-q}},
\end{equation}
for  $q<1, a,\delta,\eta > 0,\left(1-a(1-q){G(y)}^\delta \right) >0$ and $g_1(y)$ = 0 elsewhere.

\begin{equation}\label{p.2}
g_{2}(y)=(\eta + 1 - q)a\delta g(y){G(y)}^{\delta-1} \left[1+a(q-1){G(y)}^{\delta} \right]^{-\frac{\eta}{q-1}},
\end{equation}
for $q>1, a,\delta,\eta > 0, \left(1+a(q-1){G(y)}^\delta \right) >0 $ and and $g_2(y)$ = 0 elsewhere.

\begin{equation}\label{p.3}
g_{3}(y)=\eta a\delta g(y) {G(y)}^{\delta-1} e^{-a \eta {G(y)}^\delta},
\end{equation}
for $q \rightarrow 1, a>0, \delta>0, \eta>0$ and $g_3(y)$ = 0 elsewhere. 
\end{defn}

Hence in this paper we consider Type I pathway generated exponential (PGE-1) distribution, that is, in \eqref{p.1} we have $G(x) = 1-e^{-\lambda x}; x \ge 0, \lambda > 0$. The PGE-1 distribution introduced in this paper is a right truncated distribution and its applications in stress-strength reliability is studied. We are interested in this topic because we are not aware of any discussions regarding inference on $R$ when $X$ and $Y$ have a PGE-1 distribution.

 Numerous study fields naturally produce right-truncated data, such as the AIDS blood transfusion data set taken from the Centers for Disease Control (CDC) database. Chi $et ~ al.$ (2007), Lagakos $et ~ al.$ (1988), and Kalbfleisch and Lawless  (1989)   have all examined the data on AIDS blood transfusions. The data include information of transfusions cases of transfusion-related AIDS, corresponding to individuals diagnosed prior to July 1, 1986. As an example, Zaninetti (2013) provided a right and left truncated gamma distribution with application to stars. Right truncated datasets are also seen in other fields of study. Behdani $et ~ al.$ (2020) has looked into double truncated distributions with applications in income inequality. Teamah $et ~ al.$ (2020) have studied properties of right truncated Fr${\rm \grave{e}}$chet-Weibull distribution.

The rest of the paper can be summarized as follows. In section 2 type 1 pathway generated exponential distribution is introduced and its properties are discussed. In sections 3 we obtain the maximum likelihood estimate (MLE)  of $R$ and the interval estimates of $R$, where we discuss asymptotic confidence interval (ACI) and Bootstrap confidence interval (BCI). In the next section, a simulation study is conducted by using the Monte Carlo simulations to access the performance of the developed results. Also application of the model is illustrated on AIDS blood transfusion dataset. Finally, the concluding remarks are given in section 5.

\section{Type I Pathway Generated Exponential Distribution}

\begin{figure}[h] \centering
\caption{Different shapes of the density function}
\label{Den}
\includegraphics[scale=0.8]{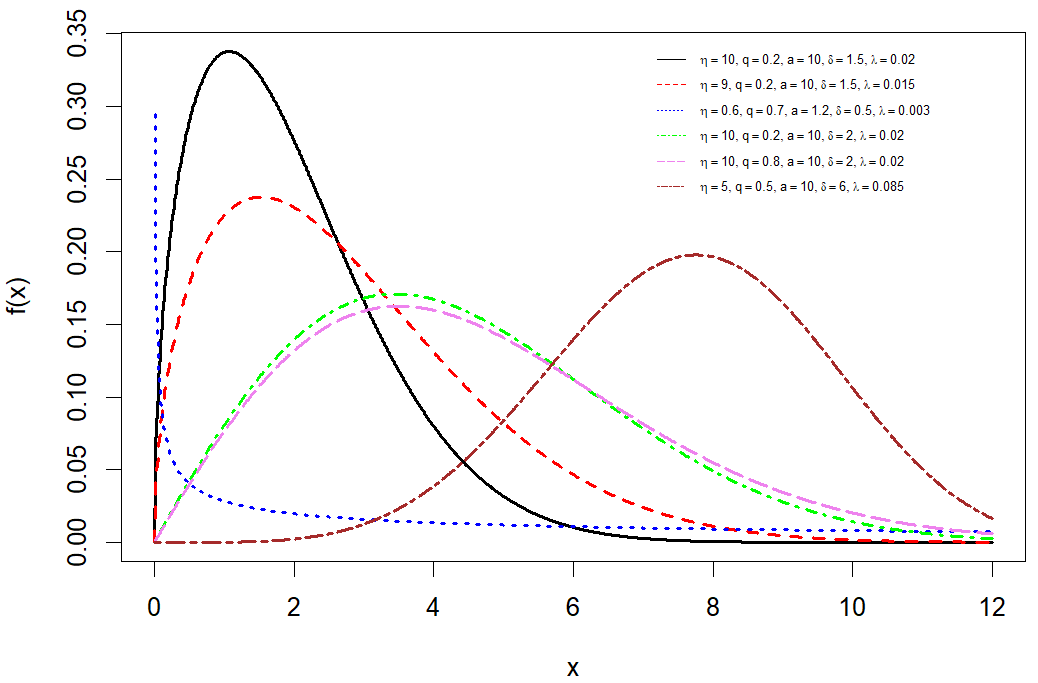}
\end{figure}

The Type I pathway generated exponential (PGE-1) distribution is obtained by taking $G(x) = 1-e^{-\lambda x}; x \ge 0, \lambda > 0$ in \eqref{p.1}.
	\begin{defn}
	 A random variable $X$ possessing the PGE-1 distribution with parameters  $(a,\delta,\lambda,\eta,q)$ has the cdf and pdf respectively are
	
 
\begin{equation}\label{pe.4}
F(x) = 1 - [1-a(1-q)(1- e^{-\lambda x})^{\delta}]^{\frac{\eta}{1-q}+1},  
 0<x< -\frac{1}{\lambda} \ln[1-(a(1-q))^{\frac{-1}{\delta}}] ,
\end{equation}

\begin{equation}\label{pe.1}
f(x) = (\eta + 1 - q) a \delta \lambda e^{-\lambda x} \left(1- e^{-\lambda x} \right)^{\delta-1} \left[1-a(1-q) \left(1- e^{-\lambda x} \right)^{\delta} \right]^{\frac{\eta}{1-q}}
\end{equation}
for $0 \le x \le -\frac{1}{\lambda} \ln\left[1-(a(1-q))^{\frac{-1}{\delta}} \right], q < 1, \delta, a, \lambda, \eta > 0, (a(1-q))^{\frac{-1}{\delta}} < 1$ and $f(x) = 0$ elsewhere. 
\end{defn}
When $\delta = 1$ and $a(1-q)=1, f(x)$ reduces to an exponential distribution with parameter $(a\eta + 1)\lambda$. The plots with different shapes of the density function are given in Figure \ref{Den}, which suggests that $\delta$ is a shape parameter, $\eta$ and $a$ are scale parameters. It can be observed that with change in $\eta$ and $a$, we can have thicker and thinner tails.

\begin{figure}[h] \centering
\caption{Different shapes of the hazard function}
\label{Hz}
 iuj\includegraphics[scale=0.8]{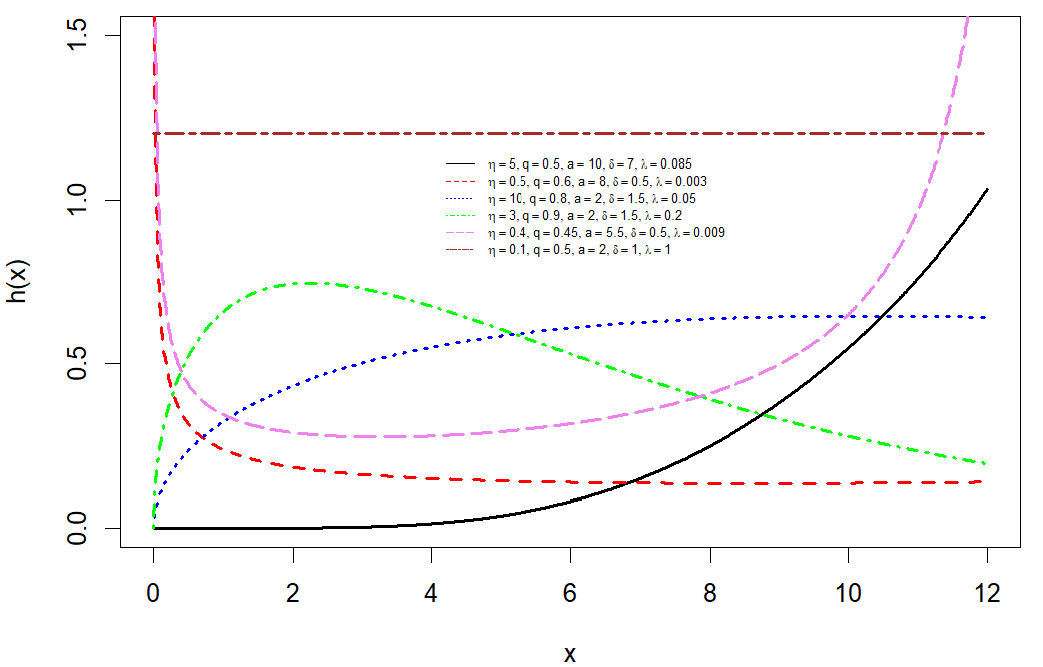}
\end{figure}

The corresponding survival function, $S(x) = Pr\{X > x\}$ and the hazard function, $h(x) = \frac{f(x)}{S(x)}$ is obtained in \eqref{Surv} and \eqref{H} respectively.

\begin{eqnarray} \label{Surv}
S(x) &=& \left[ 1-a(1-q)\left(1- e^{-\lambda x} \right)^{\delta} \right]^{\frac{\eta}{1-q}+1} \\ \label{H}
h(x) &=& \frac{(\eta + 1 - q) a \delta \lambda e^{-\lambda x} \left(1- e^{-\lambda x} \right)^{\delta-1}} {1-a(1-q) \left(1- e^{-\lambda x} \right)^{\delta}}
\end{eqnarray}
Note that $\delta = 1, a = 2$ and $q = \frac{1}{2}$, the hazard function reduces to $h(x) = (2\eta+1)\lambda,$ which is a constant hazard rate function. Figure \ref{Hz} gives different shapes of the hazard function. It can observed that a constant hazard, increasing, decreasing, monotonically increasing, bathtub and upside down bathtub shape can be attained.

\subsection{Moment generating function}

The moment generating function (mgf) is defined as for $t > 0, M_{X}(t)$
\begin{eqnarray} \nonumber
 &=& E[e^{tx}] = \int_{X} e^{tx} f(x) {\rm d}x \\ \nonumber
&=& (\eta + 1 - q) a \delta \lambda \int_{X} e^{(t-\lambda) x} \left(1- e^{-\lambda x} \right)^{\delta-1} \left[1-a(1-q) \left(1- e^{-\lambda x} \right)^{\delta} \right]^{\frac{\eta}{1-q}} {\rm d}x
\end{eqnarray}
Consider the transformation, $a(1-q)\left(1- e^{-\lambda x} \right)^{\delta} = u$, then we have
\begin{eqnarray} \nonumber
M_{X}(t) &=& (\eta + 1 - q) \int_{0}^{1} \left[ 1 - \left( \frac{u}{a(1-q)} \right)^{\frac{1}{\delta}} \right]^{\frac{-t}{\lambda}} (1-u)^{\frac{\eta}{1-q}} {\rm d}u \\ \nonumber
&=& \frac{(\eta + 1 - q)}{(a(1-q))^{\frac{-1}{\delta}}} \sum_{k=o}^{\infty} \frac{\left(\frac{t}{\lambda} \right)_{k}}{k!} \left(a(1-q)\right)^{\frac{-k}{\delta}} \int_{0}^{1} u^{\frac{k}{\delta}} (1-u)^{\frac{\eta}{1-q}} {\rm d}u \\ 
&=& \frac{(\eta + 1 - q)}{(a(1-q))^{\frac{-1}{\delta}}} \sum_{k=o}^{\infty} \frac{\left(\frac{t}{\lambda} \right)_{k}}{k!} \left(a(1-q)\right)^{\frac{-k}{\delta}} B\left( \frac{k}{\delta} + 1, \frac{\eta}{1-q} + 1 \right)
\end{eqnarray}
where $(b)_{k}$ is called the Pochhammer symbol and is defined as $(b)_{k} = b (b+1) ... (b+k-1), (b)_{0} = 1, b \neq 0$ and $B(x,y) = \int_{0}^{1} t^{x-1} (1-t)^{y-1} {\rm d}t$ is called the beta function.

\subsection{Stochastic ordering}

For two random variables $X$ and $Y$ we can say $X$ is smaller than $Y$ in likelihood ratio, denoted by, $X \le_{lr} Y$, if and only if, $\frac{f_{X}(x)}{g_{Y}(x)}$ is a decreasing function in $x$, where $f_{X}(x)$ and $g_{Y}(x)$ are density functions of $X$ and $Y$ respectively. Let  $X$ and $Y$ are independent random variables following PGE-1 distribution with parameters $(a, \delta, \lambda, \eta_{1}, q)$ and $(a, \delta, \lambda, \eta_{2}, q)$ respectively, and for $\eta_{1} \le \eta_{2}$ then
\begin{equation} \nonumber
\frac{f_{X}(x)}{g_{Y}(x)} = \left( \frac{\eta_{1}+1-q}{\eta_{2}+1-q} \right) \left[1-a(1-q) \left(1- e^{-\lambda x} \right)^{\delta} \right]^{\frac{\eta_{1}-\eta_{2}}{1-q}} 
\end{equation}
Now by differentiation with respect to $x$, we have
\begin{eqnarray} \nonumber
\frac{\partial}{\partial x}\frac{f_{X}(x)}{g_{Y}(x)} &=& a \delta \lambda \left(\eta_{1}-\eta_{2}\right) \left( \frac{\eta_{1}+1-q}{\eta_{2}+1-q} \right) e^{-\lambda x} \left(1- e^{-\lambda x} \right)^{\delta-1} \\ \nonumber
&\times& \left[1-a(1-q) \left(1- e^{-\lambda x} \right)^{\delta} \right]^{\frac{\eta_{1}-\eta_{2}}{1-q}} 
\end{eqnarray}
Clearly for $\eta_{1} \le \eta_{2}, \frac{f_{X}(x)}{g_{Y}(x)}$ is decreasing and hence, $X \le_{lr} Y$.

\subsection{Stress-strength reliability}

Suppose $X$ and $Y$ are independent random variables following PGE-1 distribution with parameters $(a, \delta, \lambda, \eta_{1}, q)$ and $(a, \delta, \lambda, \eta_{2}, q)$ respectively. Then
\begin{eqnarray} \nonumber
R &=& Pr\{X > Y\} = \int_{0}^{-\frac{1}{\lambda} \ln[1-(a(1-q))^{\frac{-1}{\delta}}]} f_{X}(x) \int_{0}^{x} g_{Y}(y) {\rm d}y {\rm d}x \\ \label{R}
&=& \frac{\eta_{2} + 1 - q}{\eta_{1} + \eta_{2} + 2(1-q)}
\end{eqnarray}

\section{Estimation of $R$}

In this section, we obtain the point and interval estimates of $R$. The point estimate of $R$ is obtained using maximum likelihood method. For interval estimates we consider asymptotic confidence interval and parametric bootstrap confidence intervals.

\subsection{Maximum Likelihood Estimation of $R$}

In this section, the maximum likelihood estimate (MLE) of the stress-strength reliability is obtained. Suppose $x_1,x_2,\dots,x_{n_1}$, is a random sample of size $n_1$ from the PGE-1 $(a, \delta, \lambda, \eta_{1}, q)$ and $y_1,y_2,\dots,y_{n_2}$ is an independent random of size $n_2$ from the PGE-1 $(a, \delta, \lambda, \eta_{2}, q)$ respectively. Then the likelihood function based on the two independent random samples is given by
\begin{eqnarray} \nonumber
L(x,y) &=& \prod_{i=1}^{n_1} f_X(x;a,\delta,\lambda,\eta_1,q)\prod_{j=1}^{n_2} g_Y(y;a,\delta,\lambda,\eta_2,q)  
\end{eqnarray}

Then the log-likelihood function of the observed sample is
\begin{eqnarray} \nonumber
\ell &=& n_1 \ln(\eta_1 +1-q) +n_2 \ln(\eta_2 +1-q) +(n_1+n_2) \ln(\alpha\delta\lambda) \\ \nonumber 
&-& \lambda \left(\sum_{i=1}^{n_{1}} x_{i} + \sum_{j=1}^{n_{2}} y_{j}\right) + (\delta-1)  \left[ \sum_{i=1}^{n_1} \ln\left(1-e^{-\lambda x_i}\right) + \sum_{j=1}^{n_2} \ln\left(1-e^{-\lambda y_j}\right) \right] \\ \nonumber
&+& \frac{\eta_1}{1-q} \left[ \sum_{i=1}^{n_1} \ln \left(1-a(1-q)\left(1-e^{-\lambda x_i}\right)^{\delta} \right) \right] \\
&+& \frac{\eta_2}{1-q} \left[  \sum_{j=1}^{n_1} \ln \left(1-a(1-q)\left(1-e^{-\lambda y_j}\right)^{\delta} \right) \right]
\end{eqnarray}

The MLEs of $a, \delta, \lambda, \eta_1, \eta_2$ and $q$, say $\hat{a},\hat{\delta},\hat{\lambda},\hat{\eta_1},\hat{\eta_2}$ and $\hat{q}$ respectively can be obtained as the solution of 
\begin{equation*}
\frac{\partial \ell}{\partial a} =0; ~ \frac{\partial \ell}{\partial \delta} =0; ~ \frac{\partial \ell}{\partial \eta_1} =0; ~ \frac{\partial \ell}{\partial \eta_2} =0; ~ \frac{\partial \ell}{\partial \lambda} =0; ~ \frac{\partial \ell}{\partial q} =0
\end{equation*}

Where the first order partial derivatives are given by,
\begin{eqnarray*}
\frac{\partial \ell}{\partial a} &=&  \frac{n_1+n_2}{a} -\eta_1 \sum_{1=1}^{n_1} \left[\frac{(1-e^{-\lambda x_i})^\delta}{1-a(1-q)(1-e^{-\lambda x_i})^\delta} \right] \\ 
&-& \eta_2 \sum_{j=1}^{n_2} \left[ \frac{(1-e^{-\lambda y_j})^\delta}{1-a(1-q)(1-e^{-\lambda y_j})^\delta} \right]   \\
\frac{\partial \ell}{\partial \delta} &=& \frac{n_1+n_2}{\delta} + \sum_{i=1}^{n_1} \ln(1-e^{-\lambda x_i}) + \sum_{j=1}^{n_2} \ln(1-e^{-\lambda y_j}) \\ 
&-& \eta_1 \sum_{i=1}^{n_1} \left[\frac{a(1-e^{-\lambda x_i})^\delta \ln (1-e^{-\lambda x_i})}{1-a(1-q)(1-e^{-\lambda x_i})^\delta}  \right] - \eta_2 \sum_{j=1}^{n_2} \left[ \frac{a(1-e^{-\lambda y_j})^\delta \ln (1-e^{-\lambda y_j})}{1-a(1-q)(1-e^{-\lambda y_j})^\delta} \right]   
\end{eqnarray*}
\begin{eqnarray*}
\frac{\partial \ell}{\partial \eta_1}  &=& \frac{n_1}{\eta_1+1-q} + \frac{1}{1-q} \left[ \sum_{i=1}^{n_1} \ln \left(1-a(1-q)\left(1-e^{-\lambda x_i}\right)^{\delta} \right) \right]  \\
\frac{\partial \ell}{\partial \eta_2}  &=& \frac{n_2}{\eta_2+1-q} + \frac{1}{1-q} \left[ \sum_{j=1}^{n_2} \ln \left(1-a(1-q)\left(1-e^{-\lambda y_j}\right)^{\delta} \right) \right]  \\
\frac{\partial \ell}{\partial \lambda}  &=& \frac{n_1+n_2}{\lambda} - \left(\sum_{i=1}^{n_{1}} x_{i} + \sum_{j=1}^{n_{2}} y_{j}\right) + (\delta-1) \left[ \sum_{i=1}^{n_1} \frac{x_i e^{-\lambda x_i}}{1-e^{-\lambda x_i}} + \sum_{j=1}^{n_2} \frac{y_j e^{-\lambda y_j}}{1-e^{-\lambda y_j}} \right] \\ 
&-& \sum_{i=1}^{n_1} \frac{\eta_1 \delta a x_i e^{-\lambda x_i} (1-e^{-\lambda x_i})^{\delta-1}} {1-a(1-q)(1-e^{-\lambda x_i})^\delta} - \sum_{j=1}^{n_2} \frac{\eta_2 \delta a y_j e^{-\lambda y_j} (1-e^{-\lambda y_j})^{\delta-1}} {1-a(1-q)(1-e^{-\lambda y_j})^\delta} \\
\frac{\partial \ell}{\partial q} &=& \frac{n_1}{\eta_1+1-q}+ \frac{n_2}{\eta_2+1-q} + \frac{\eta_1}{(1-q)^{2}} \left[ \sum_{i=1}^{n_1} \ln \left(1-a(1-q)\left(1-e^{-\lambda x_i}\right)^{\delta} \right) \right] \\
&+& \frac{\eta_2}{(1-q)^{2}} \left[ \sum_{j=1}^{n_2} \ln \left(1-a(1-q)\left(1-e^{-\lambda y_j}\right)^{\delta} \right) \right] \\
&+& \frac{\eta_1}{1-q} \sum_{i=1}^{n_1} \left[\frac{a (1-e^{-\lambda x_i})^\delta}{1-a(1-q)(1-e^{-\lambda x_i})^\delta} \right] + \frac{\eta_2}{1-q} \sum_{j=1}^{n_2} \left[\frac{a (1-e^{-\lambda y_j})^\delta}{1-a(1-q)(1-e^{-\lambda y_j})^\delta} \right]
\end{eqnarray*}
The MLEs of the parameters can be obtained by solving the nonlinear likelihood equations using, for example, the Newton-Raphson iteration scheme. Because of the higher number of parameters, it is suggested that careful consideration should be provided while selecting the initial values. The corresponding MLE of $R$ is computed from \eqref{R} after replacing $\eta_1,\eta_2$ and $q$ by their MLEs. Thus we have the estimate of stress-strength reliability as
\begin{equation}
\hat{R}=\frac{\hat{\eta_2} +1 -\hat{q}}{\hat{\eta_1} +\hat{\eta_2}+2(1-\hat{q})}
\end{equation}

\subsection{Asymptotic confidence interval for $R$}

As it is difficult to compute the exact distribution of $R$, we obtain the asymptotic confidence interval. Let us consider the Fisher information matrix, 
$$
I=-E\left[ \begin{array}{cccccc}
\frac{\partial^{2} \ell}{\partial a^{2}} & \frac{\partial^{2} \ell}{\partial a \partial \delta} & \frac{\partial^{2} \ell}{\partial a \partial \eta_{1}} & \frac{\partial^{2} \ell}{\partial a \partial \eta_{2}} & \frac{\partial^{2} \ell}{\partial a \partial \lambda} & \frac{\partial^{2} \ell}{\partial a \partial q} \\
\frac{\partial^{2} \ell}{\partial \delta \partial a} & \frac{\partial^{2} \ell}{\partial \delta^{2}} & \frac{\partial^{2} \ell}{\partial \delta \partial \eta_{1}} & \frac{\partial^{2} \ell}{\partial \delta \partial \eta_{2}} & \frac{\partial^{2} \ell}{\partial \delta \partial \lambda} & \frac{\partial^{2} \ell}{\partial \delta\partial q} \\	
\frac{\partial^{2} \ell}{\partial \eta_{1} \partial a} & \frac{\partial^{2} \ell}{\partial \eta_{1} \partial \delta} & \frac{\partial^{2} \ell}{\partial \eta_{1}^{2}} & \frac{\partial^{2} \ell}{\partial \eta_{1} \partial \eta_{2}} & \frac{\partial^{2} \ell}{\partial \eta_{1} \partial \lambda} & \frac{\partial^{2} \ell}{\partial \eta_{1} \partial q} \\
\frac{\partial^{2} \ell}{\partial \eta_{2} \partial a} & \frac{\partial^{2} \ell}{\partial \eta_{2} \partial \delta} & \frac{\partial^{2} \ell}{\partial \eta_{2} \partial \eta_{1}} & \frac{\partial^{2} \ell}{\partial \eta_{2}^{2}} & \frac{\partial^{2} \ell}{ \partial \eta_{2} \partial \lambda} & \frac{\partial^{2} \ell}{ \partial \eta_{2} \partial q} \\
\frac{\partial^{2} \ell}{\partial \lambda \partial a} & \frac{\partial^{2} \ell}{\partial \lambda  \partial \delta} & \frac{\partial^{2} \ell}{\partial \lambda  \partial \eta_{1}} & \frac{\partial^{2} \ell}{\partial \lambda \partial \eta_{2} } & \frac{\partial^{2} \ell}{\partial \lambda^{2}} & \frac{\partial^{2} \ell}{\partial \lambda  \partial q} \\
\frac{\partial^{2} \ell}{\partial q \partial a} & \frac{\partial^{2} \ell}{\partial q  \partial \delta} & \frac{\partial^{2} \ell}{\partial q  \partial \eta_{1}} & \frac{\partial^{2} \ell}{\partial q \partial \eta_{2} } & \frac{\partial^{2} \ell}{\partial q\partial \lambda} & \frac{\partial^{2} \ell}{\partial q^{2}}
\end{array}\right]
$$
The second order partial derivatives are given in the appendix. Due to the complexity of the expectations, under mild conditions, the observed information matrix can be used as a consistent estimator of the information matrix, $I$. An approximate estimate of the variance-covariance matrix of $(a,\delta,\lambda,\eta_1,\eta_2,q)$ is $I^{-1}$.

Let $G^{T} = \left(\frac{\partial R}{\partial a},\frac{\partial R}{\partial \delta},\frac{\partial R}{\partial \eta_1},\frac{\partial R}{\partial \eta_2},\frac{\partial R}{\partial \lambda},\frac{\partial R}{\partial q} \right)$, where
\begin{eqnarray*}
\frac{\partial R}{\partial \eta_1} &=&  -\frac{\eta_2 + 1 - q}{(\eta_1 + \eta_2 + 2(1-q))^2} \\
\frac{\partial R}{\partial \eta_2} &=& \frac{\eta_1 + 1 - q}{(\eta_1 + \eta_2 + 2(1 - q))^2} \\
\frac{\partial R}{\partial q} &=&  \frac{\eta_{1}-\eta_{2}}{(\eta_1 + \eta_2 + 2(1 - q))^2} 
\end{eqnarray*}
As $R$ does not depend on $a, \delta$ and $\lambda$, we have $ \frac{\partial R}{\partial a} = \frac{\partial R}{\partial \delta} = \frac{\partial R}{\partial \lambda} = 0$.

Then, the approximate estimate of Var($\hat{R}$) is $\hat{Var}(\hat{R}) \approx G^{T}I^{-1}G$. Thus, $\frac{\hat{R}-R}{\sqrt{\hat{Var}(\hat{R})}} \sim N(0,1)$ asymptotically. This result yields the asymptotic $100(1-\gamma) \%$ confidence interval for $R$ as 
$\hat{R} \pm Z_{\frac{\gamma}{2}} \sqrt{\hat{Var}(\hat{R})}$
where $Z_{\frac{\gamma}{2}}$ is the upper $\frac{\gamma}{2}$ percentile of the standard normal distribution.

\subsection{Bootstrap confidence interval for $R$}

It is anticipated that confidence intervals based on asymptotic results will not perform well for small sample sizes (s). In this subsection,   we suggest using two confidence intervals based on parametric bootstrap methods: (i)   the percentile bootstrap technique (Boot-$p$), which is based on Efron's (1982) theory, and (ii) the bootstrap-t method (Boot-$t$), which is based on Hall's  (1988)   theory.  The algorithms for estimating the confidence intervals of $R$ using both methods are demonstrated below.\\

\noindent \textbf{(i) Boot-p Method} \\

\noindent \textbf{Step 1:} Based on the independent samples $\underline{x}=\left\{x_{1}, \ldots, x_{n_1}\right\}$ and $\underline{y}=\left\{y_{1}, \ldots, y_{n_2}\right\}$ from the PGE-1 $(a, \delta, \eta_1, \lambda, q)$ and PGE-1 $(a, \delta, \eta_2, \lambda, q)$, respectively. Compute the MLE $(\hat{a}, \hat{\delta}, \hat{\lambda}, \hat{\eta_1}, \hat{\eta_2}, \hat{q})$ of $(a, \delta, \eta_1, \eta_2, \lambda, q)$.\\
\textbf{Step 2:} Using $\hat{a}, \hat{\delta}, \hat{\lambda}, \hat{\eta_1}, \hat{q}$, generate an independent parametric bootstrap sample $\left\{x_{1}^{*}, \ldots, x_{n_1}^{*}\right\}$, and similarly using $\hat{a}, \hat{\delta}, \hat{\lambda}, \hat{\eta_2}, \hat{q}$, generate a bootstrap sample $\left\{y_{1}^{*}, \ldots, y_{n_2}^{*}\right\}$ . Use these samples to compute the MLE $\hat{a}^{*}, \hat{\delta}^{*}, \hat{\lambda}^{*}, \hat{\eta_1}^{*}, \hat{\eta_2}^{*}, \hat{q}^{*}$, and compute the bootstrap estimate of $R$ using Equation \eqref{R}, say $\hat{R}^{*}$.\\
\textbf{Step 3:} Repeat step 2, $N$ times to get the parametric bootstrap estimates $\hat{R}_{1}^{*}, \ldots, \hat{R}_{N}^{*}$ of $R$.\\
\textbf{Step 4:} Let $H_1(x)=P(\hat{R} \leq x)$ be the cumulative distribution function of $\hat{R}$. Define $\hat{R}_{BP}(x)=$ $H_1^{-1}(x)$ for a given $x$. The approximate $100(1-\gamma) \%$ confidence interval of $R$ is given by $$\left(\hat{R}_{BP}\left(\frac{\gamma}{2}\right), \hat{R}_{BP}\left(1-\frac{\gamma}{2}\right) \right)$$. \\

\noindent \textbf{(ii) Boot-t Method} \\

\noindent \textbf{Step 1:} Based on the independent samples $\underline{x}=\left\{x_{1}, \ldots, x_{n_1}\right\}$ and $\underline{y}=\left\{y_{1}, \ldots, y_{n_2}\right\}$ from the PGE-1 $(a, \delta, \eta_1, \lambda, q)$ and PGE-1 $(a, \delta, \eta_2, \lambda, q)$, respectively. Compute the MLE $(\hat{a}, \hat{\delta}, \hat{\lambda}, \hat{\eta_1}, \hat{\eta_2}, \hat{q})$ of $(a, \delta, \eta_1, \eta_2, \lambda, q)$.\\
\textbf{Step 2:} Using $\hat{a},\hat{\delta},\hat{\lambda},\hat{\eta_1},\hat{q}$, generate an independent parametric bootstrap sample $\left\{x_{1}^{*}, \ldots, x_{n_1}^{*}\right\}$, and similarly using $\hat{a},\hat{\delta},\hat{\lambda},\hat{\eta_2},\hat{q}$, generate a bootstrap sample $\left\{y_{1}^{*}, \ldots, y_{n_2}^{*}\right\}$ . Use these samples to compute the MLE $\hat{a}^{*},\hat{\delta}^{*},\hat{\lambda}^{*},\hat{\eta_1}^{*},\hat{\eta_2}^{*},\hat{q}^{*}$, and compute the bootstrap estimate of $R$ using Equation \eqref{R}, say $\hat{R}^{*}$ and the statistic,
$$
T^{*}=\frac{\sqrt{n}\left(\widehat{R}^{*}-\widehat{R}\right)}{\sqrt{\operatorname{Var}\left(\widehat{R}^{*}\right)}} .
$$
Here variance can be calculated as discussed in section 3.2. \\
\textbf{Step 3:} Repeat step 2, $N$ times to get the parametric bootstrap estimates $\hat{R}_{1}^{*}, \ldots, \hat{R}_{N}^{*}$ of $R$.\\
\textbf{Step 4:} Let $H_{2}(x)=P\left(T^{*} \leq x\right)$ be the cumulative distribution function of $T^{*}$. For a given $x$, define $\widehat{R}_{B T}(x)=\widehat{R}+H_{2}^{-1}(x) \sqrt{\frac{\operatorname{Var}(\widehat{R})}{n}}$. The approximate $100(1-\gamma) \%$ confidence interval of $R$ is given by
$$\left(\widehat{R}_{B t}\left(\frac{\gamma}{2}\right), \widehat{R}_{B t}\left(1-\frac{\gamma}{2}\right)\right).$$

\section{Simulation study and data analysis}

In this section simulation study is considered to verify the performance of the developed methods. Also the application of the results is illustrated on
AIDS incubation times.

\subsection{Simulation study}

Here we illustrate the performance of the maximum likelihood estimate (MLE) of $R$, the performance is measured in terms of biases and mean square error (MSE). The performance of the confidence intervals is measured in terms of confidence length (CL). Random samples of sizes $(n_1,n_2) = n = \{$(10,10), (20,20), (30,30), (50,50) and (100, 100)$\}$ is considered. We consider three different value of $R = \{$0.1428 $(\alpha = 4, \delta = 1,\lambda = 1.5, \eta_1 = 10, \eta_2 = 1, q = 0.2)$, 0.5362 $(\alpha = 1, \delta = 1, \lambda = 0.5, \eta_1 = 2.5, \eta_2 = 3, q = 0.3)$, 0.9054 $(\alpha = 10, \delta = 1, \lambda = 0.5, \eta_1 = 2, \eta_2 = 20, q = 0.9) \}$.
\begin{table}[H]
\caption{Performance of the developed methods}
\centering
\begin{tabular}{c@{\hskip 0.1in}c@{\hskip 0.1in}c@{\hskip 0.1in}c@{\hskip 0.1in}c@{\hskip 0.1in}c@{\hskip 0.1in}c@{\hskip 0.1in}}
\noalign{\smallskip} \hline \noalign{\smallskip}
n & $\hat{R}$ & Bias & MSE & CI & Boot-p CI & Boot-t CI \\ \noalign{\smallskip}
\hline \noalign{\smallskip}
10 & 0.1713 & 0.0284 & 0.0068 & (0.1255, 0.2124) & (0.0558, 0.3390) & (0.0399, 0.3635) \\
20 & 0.1638 & 0.0210 & 0.0055 & (0.1245, 0.2020) & (0.0486, 0.3140) & (0.0624, 0.3529) \\
30 & 0.1418 & -0.0010 & 0.0013 & (0.1222, 0.1614) & (0.0820, 0.2192) & (0.0654, 0.2919) \\
50 & 0.1408 & -0.0020 & 0.0008 & (0.1259, 0.1558) &   (0.0904, 0.1996) & (0.0708, 0.2895) \\
100 & 0.1416 & -0.0012 & 0.0004 & (0.1309, 0.1523) &  (0.1031, 0.1831) & (0.1412, 0.1834) \\ \noalign{\smallskip}
\hline \noalign{\smallskip}
10 & 0.5659 & 0.0297 & 0.0019 & (0.5241, 0.6078) & (0.5250, 0.5959) & (0.5243, 0.6088) \\
20 & 0.5379 & 0.0017 & 0.0008 & (0.4979, 0.5780) & (0.3553, 0.7045) & (0.4872, 0.6672) \\
30 & 0.5159 & -0.0203 & 0.0005 & (0.4720, 0.5598) & (0.4659, 0.5559) &  (0.4720, 0.5598) \\
50 & 0.5359 & -0.0003 & 0.0002 & (0.5100, 0.5617) & (0.4852, 0.5616) &  (0.4774, 0.5863) \\
100 & 0.5381 & 0.0019 & 0.0001 & (0.5201,0.5561) & (0.4937, 0.5586) & (0.5010, 0.5574) \\ \noalign{\smallskip}
\hline \noalign{\smallskip}
10 & 0.8925 & -0.0129 & 0.0032 & (0.8624, 0.9223) & (0.7682, 0.9710) &  (0.7960, 0.9980) \\
20 & 0.9042 & -0.0012 & 0.0015 & (0.8856, 0.9227) & (0.8207, 0.9586) &  (0.8674, 0.9782) \\
30 & 0.8977 & -0.0077 & 0.0013 & (0.8812, 0.9141) & (0.8967, 0.9154) &  (0.8814, 0.9138) \\
50 & 0.9087 & 0.0033 & 0.0005 & (0.8972, 0.9202) & (0.8990, 0.9159) & (0.8840, 0.9107) \\
100 & 0.9072 & 0.0018 & 0.0003 & (0.8991, 0.9152) & (0.8944, 0.9161) & (0.8960, 0.9189) \\ \noalign{\smallskip}
\hline \noalign{\smallskip}
\end{tabular}
\label{tab:my_label}
\end{table}

\subsection{Example - AIDS blood transfusion data}

The AIDS blood transfusion data is collected by Center for Disease Control (CDC) that administrates a national registry of AIDS patients. Here the event time is the latent period of the acquired immune deficiency syndrome (AIDS) which is defined as the time from HIV virus infection to the diagnosis of AIDS. As the transfusion of the contaminated blood is a source of the virus, if one receives blood and later was diagnosed of AIDS, then the time patient was infected can be traced. Here in this study, a subject is included in the sample only when the event of interest, AIDS diagnosis, occurs before the closing date of the study. Thus the AIDS latent time is being right truncated.

The data includes 295 cases, and diagnosed prior to July 1, 1986, and can be obtained from Kalbfleisch and Lawless (1989), and is also available in {\it SurvTrunc} package in R. The data also record the age of the patients. The latent time is measured in months, there was no observation with value 0 and it has been excluded from the analysis. We divide the data into groups - age less than or equal to 16 (teens) and age greater than 16 (adults). Hence we consider $X$ as the latent time for teens and $Y$ as the latent time for adults. 

\begin{table}[h]
\begin{center}
\caption{Goodness of fit for data set 1} \label{Fit 1} 
\begin{tabular}{cccccccc}
\hline  \noalign{\smallskip}
Data set & $a$ & $\delta$ & $\eta$ & $\lambda$ & $q$ & K-S Statistic & $p$-value \\
\hline  \noalign{\smallskip}
$X$ & 10.5 & 1.9 & 10.5 & 0.0040 & 0.5 & 0.1284 & 0.5758 \\
$Y$ & 10.5 & 1.9 & 3.0 & 0.0040 & 0.5 & 0.0519 & 0.4935 \\
\hline
\end{tabular}
\end{center}
\end{table}

\begin{figure}[h] \centering
\caption{Histogram and fitted pdf}
\label{H1}
\begin{tabular}{cc}
\includegraphics[scale=0.4]{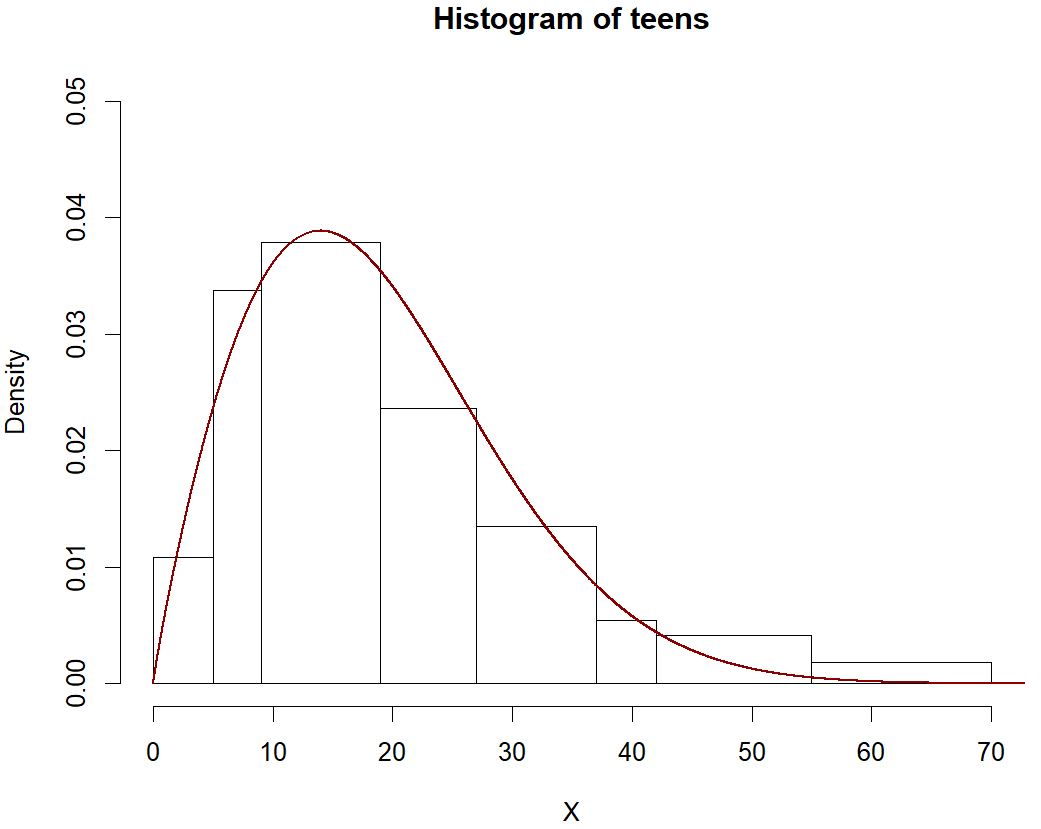} & \includegraphics[scale=0.4]{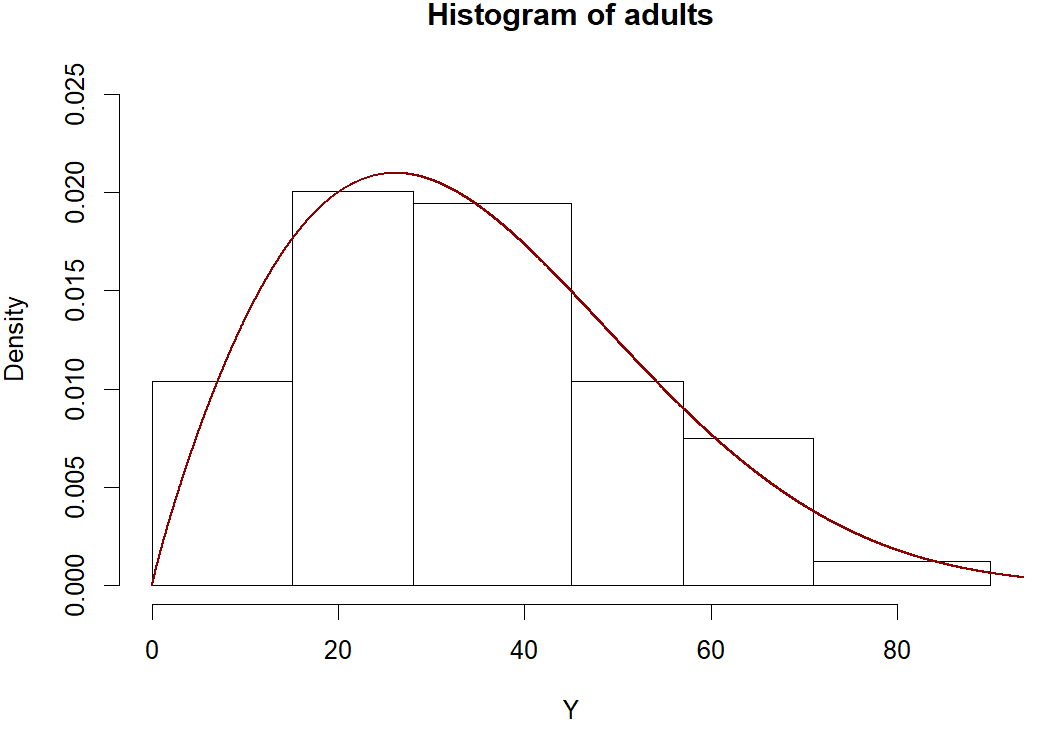}
\end{tabular}
\end{figure}

\begin{figure}[h] \centering
\caption{QQ plot with fits}
\label{QQ1}
\begin{tabular}{cc}
\includegraphics[scale=0.4]{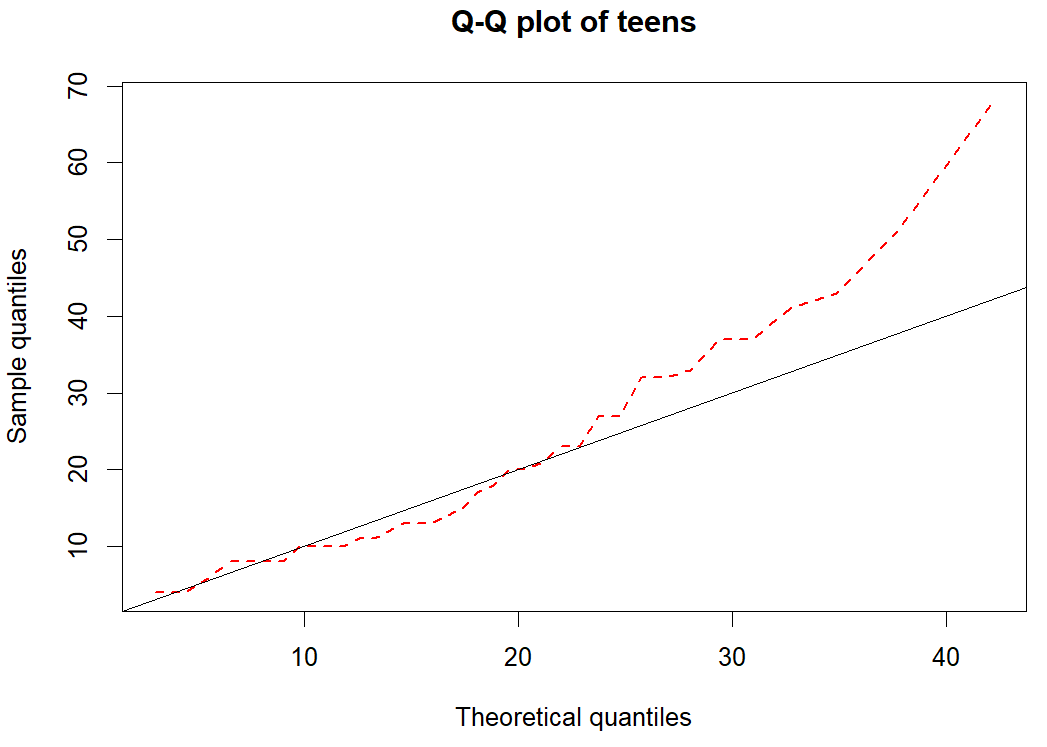} & \includegraphics[scale=0.4]{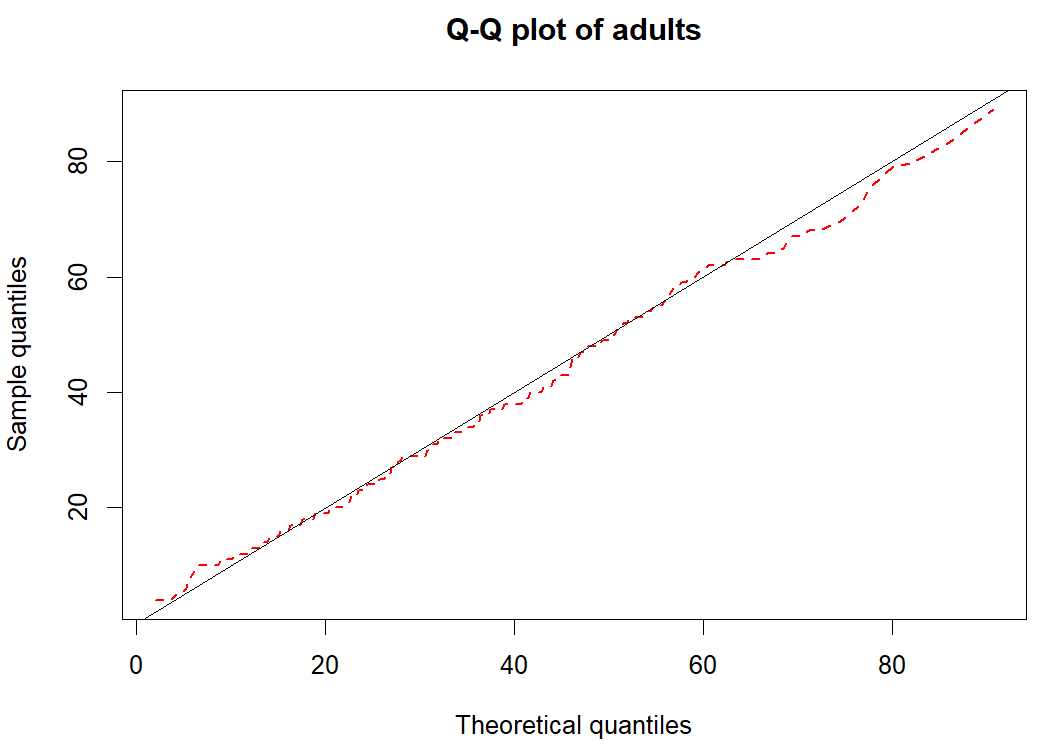}
\end{tabular}
\end{figure}

We check whether the PGE-1 distribution fits both $X$ and $Y$. The Table \ref{Fit 1}  gives the MLEs of the unknown parameters for both $X$ and $Y$. The Kolmogrov-Smirnov (K-S) test statistic and the corresponding $p$-values are also given which suggests that one cannot reject the hypothesis that the data are coming from PGE-1 distribution. Thus we have the estimate of $R$ as 0.2414 which suggests that the probability that the latent time for teens is greater than adults is 0.2414. The histogram and Q-Q plot for fits are given in Figure \ref{H1} and \ref{QQ1} respectively.

\section{Conclusions}

In this paper, we study the maximum likelihood estimation of the stress-strength reliability parameter based on Type-1 pathway generated exponential distribution. Interval estimates are also obtained using asymptotic property and bootstrap sampling methods. The application of the developed results is illustrated on blood transfusion data. The incubation time is compared among teens (age less than or equal to 16) and adults (age greater than 16) and the probability is obtained as 0.2414.

\section*{Appendix}

The diagonal elements of the matrix $I$ are:
\begin{eqnarray*}
\frac{\partial \ell^2}{\partial a^2} &=& -\frac{n_1+n_2}{a^2} + \eta_1 (1-q) \sum_{i=1}^{n_1} \frac{(1-e^{-\lambda x_i})^{2\delta}}{(1-a(1-q)(1-e^{-\lambda x_i})^\delta)^2} \\ 
&+& \eta_2 (1-q) \sum_{j=1}^{n_2} \frac{(1-e^{-\lambda y_j})^{2\delta}}{(1-a(1-q)(1-e^{-\lambda y_j})^\delta)^2}   \\
\frac{\partial \ell^2}{\partial \delta^2} &=& -\frac{(n_1+n_2)^2}{\delta^2} -\eta_1 \sum_{i=1}^{n_1} \frac{a(1-e^{-\lambda x_j})^\delta (\ln (1-e^{-\lambda x_j}))^2}{(1-a(1-q)(1-e^{-\lambda x_j})^\delta)^2} \\ 
&-& \eta_2 \sum_{j=1}^{n_2} \frac{a(1-e^{-\lambda y_j})^\delta (\ln (1-e^{-\lambda y_j}))^2}{(1-a(1-q)(1-e^{-\lambda y_j})^\delta)^2} \\
\frac{\partial \ell^2}{\partial \eta_1^2} &=& - \frac{n_1}{(\eta_1+1-q)^2}; ~~~ \frac{\partial \ell^2}{\partial \eta_2^2} = -\frac{n_2}{(\eta_2+1-q)^2} \\
\frac{\partial \ell^2}{\partial \lambda^2} &=& (\delta-1) \left[ \sum_{i=1}^{n_1} \frac{x_i^{2}e^{-2\lambda x_i} + x_i^{2}e^{-2\lambda x_i} (1-e^{-2\lambda x_i})}{(1-e^{-\lambda x_i})^2} \right. \\
&+& \left. \sum_{j=1}^{n_2} \frac{y_j^{2}e^{-2\lambda y_j} + y_j^{2}e^{-2\lambda y_j} (1-e^{-2\lambda y_j})}{(1-e^{-\lambda y_j})^2}  \right] -\frac{n_1+n_2}{\lambda^2} \\
&+& \sum_{i=1}^{n_1} \left[ \frac{\eta_1 a^2 x_i^{2}\delta e^{-2\lambda x_i} (1-q) + (1-a(1-q)(1-e^{-\lambda x_i})^\delta) a \delta x_i^{2}e^{-2\lambda x_i} (1-e^{-\lambda x_i}) }{(1-a(1-q)(1-e^{-\lambda x_i})^\delta)^2}  \right] \\ 
&+& \sum_{j=1}^{n_2} \left[ \frac{\eta_2 a^2 y_j^{2}\delta e^{-2\lambda y_j} (1-q) + (1-a(1-q)(1-e^{-\lambda y_j})^\delta) a \delta y_j^{2}e^{-2\lambda y_j} (1-e^{-\lambda y_j}) }{(1-a(1-q)(1-e^{-\lambda y_j})^\delta)^2}  \right] \\
\frac{\partial \ell^2}{ \partial q^2} &=& \frac{n_1}{(\eta_1+1-q)^2}+ \frac{n_2}{(\eta_2+1-q)^2} + \frac{2}{(1-q)^3} \sum_{i=1}^{n_1} \eta_1 \ln (1-a(1-q)(1-e^{-\lambda x_i})^\delta) \\ 
&+& \frac{2}{(1-q)^3} \sum_{j=1}^{n_2} \eta_2 \ln \left(1-a(1-q)(1-e^{-\lambda y_j})^\delta \right) \\
&+& \frac{1}{(1-q)^2} \left[ \sum_{i=1}^{n_1} \frac{\eta_1 a (1-e^{-\lambda x_i})^{\delta}}{1-a(1-q)(1-e^{-\lambda x_i})^\delta} + \sum_{j=1}^{n_2}  \frac{ \eta_2 a (1-e^{-\lambda y_j})^{\delta}}{1-a(1-q)(1-e^{-\lambda y_j})^\delta} \right] \\
&-& \frac{1}{1-q} \left[ \sum_{i=1}^{n_1} \frac{\eta_1 a^2 (1-e^{-\lambda x_i})^{2 \delta}}{(1-a(1-q)(1-e^{-\lambda x_i})^\delta)^2} + \sum_{j=1}^{n_2} \frac{ \eta_2 a^2 (1-e^{-\lambda y_j})^{2 \delta} }{(1-a(1-q)(1-e^{-\lambda y_j})^\delta)^2} \right]
\end{eqnarray*}
The non-diagonal elements of the matrix $I$ are given by:

\begin{eqnarray*}
\frac{\partial \ell^2}{\partial a \partial \delta } &=& \eta_1 \sum_{i=1}^{n_1} \left[ \frac{a(1-q)(1-e^{-\lambda x_i})^{2\delta} \ln (1-e^{-\lambda x_i})}{(1-a(1-q)(1-e^{-\lambda x_i})^\delta)^2} + \frac{(1-e^{-\lambda x_i})^{\delta} \ln (1-e^{-\lambda x_i})}{1-a(1-q)(1-e^{-\lambda x_i})^\delta} \right] \\ 
&+&  \eta_2 \sum_{j=1}^{n_2} \left[ \frac{a(1-q)(1-e^{-\lambda y_j})^{2\delta} \ln (1-e^{-\lambda y_j})}{(1-a(1-q)(1-e^{-\lambda y_j})^\delta)^2} + \frac{(1-e^{-\lambda y_j})^{\delta} \ln (1-e^{-\lambda y_j})}{1-a(1-q)(1-e^{-\lambda y_j})^\delta} \right] \\
\frac{\partial \ell^2}{\partial a \partial \eta_1 } &=& \sum_{i=1}^{n_1} \frac{-(1-e^{-\lambda x_i})^{\delta}}{1-a(1-q)(1-e^{-\lambda x_i})^\delta} ; ~~~  \frac{\partial \ell^2}{\partial a \partial \eta_2 } = \sum_{i=1}^{n_2} \frac{-(1-e^{-\lambda y_i})^{\delta} }{1-a(1-q)(1-e^{-\lambda y_i})^\delta} \\
\frac{\partial \ell^2}{\partial a \partial \lambda} &=&  \sum_{i=1}^{n_1} \frac{\eta_1 \delta  x_i e^{-\lambda x_i} (1-e^{-\lambda x_i})^{\delta-1}} {1-a(1-q)(1-e^{-\lambda x_i})^\delta)^2} + \sum_{j=1}^{n_2} \frac{\eta_2 \delta  y_j e^{-\lambda y_j} (1-e^{-\lambda y_j})^{\delta-1}} {(1-a(1-q)(1-e^{-\lambda y_j})^\delta)^2} \\
\frac{\partial \ell^2}{\partial a \partial q} &=& \sum_{i=1}^{n_1} \frac{ \eta_1 a (1-e^{-\lambda x_i})^{2\delta}}{(1-a(1-q)(1-e^{-\lambda x_i})^\delta)^2} + \sum_{j=1}^{n_2} \frac{\eta_2 a (1-e^{-\lambda y_j})^{2\delta}}{(1-a(1-q)(1-e^{-\lambda y_j})^\delta)^2} \\
\frac{\partial \ell^2}{\partial \delta \partial \eta_1} &=& \sum_{i=1}^{n_1} \frac{-a (1-e^{-\lambda x_i})^\delta \ln (1-e^{-\lambda x_i})}{1-a(1-q)(1-e^{-\lambda x_i})^\delta}; ~~~  \frac{\partial \ell^2}{\partial \delta \partial \eta_2} = \sum_{i=1}^{n_2} \frac{-a (1-e^{-\lambda y_j})^\delta \ln (1-e^{-\lambda y_j})}{1-a(1-q)(1-e^{-\lambda y_j})^\delta} \\
\frac{\partial \ell^2}{\partial \delta \partial \lambda} &=&   \sum_{j=1}^{n_2} \left[ \frac{x_i e^{-\lambda x_i}}{1-e^{-\lambda x_i}} + \frac{\eta_1 a \ln (1-e^{-\lambda x_i}) + (1-a(1-q)(1-e^{-\lambda x_i})^\delta) a x_i e^{-\lambda x_i} }{(1-a(1-q)(1-e^{-\lambda x_i})^\delta)^2} \right] \\ 
&+& \sum_{j=1}^{n_2} \left[ \frac{y_j e^{-\lambda y_j}}{1-e^{-\lambda y_j}} + \frac{\eta_2 a \ln (1-e^{-\lambda y_j}) + (1-a(1-q)(1-e^{-\lambda y_j})^\delta) a y_j e^{-\lambda y_j} }{(1-a(1-q)(1-e^{-\lambda y_j})^\delta)^2} \right] \\
\frac{\partial \ell^2}{\partial \delta \partial q} &=& - \sum_{i=1}^{n_1}  \frac{\eta_1 a^2 (1-e^{-\lambda x_i})^{2\delta} \ln (1-e^{-\lambda x_i}) }{(1-a(1-q)(1-e^{-\lambda x_i})^\delta)^2} - \sum_{j=1}^{n_2} \frac{ \eta_2 a^2 (1-e^{-\lambda y_j})^{2\delta} \ln (1-e^{-\lambda x_i}) }{(1-a(1-q)(1-e^{-\lambda y_j})^\delta)^2} \\
\frac{\partial \ell^2}{\partial \eta_1 \partial \eta_2}  &=& 0; ~~~  \frac{\partial \ell^2}{\partial \eta_1 \partial \lambda}  = - \sum_{i=1}^{n_1} \frac{a \delta  x_i e^{-\lambda x_i} (1-e^{-\lambda x_i})^{\delta-1}} {1-a(1-q)(1-e^{-\lambda x_i})^\delta} \\
\frac{\partial \ell^2}{\partial \eta_1 \partial q} &=& \frac{n_1}{(\eta_1+1-q)^2} - \sum_{i=1}^{n_1} \frac{\ln (1-a(1-q)(1-e^{-\lambda x_i})^\delta)}{(1-q)^2} \\
&+& \sum_{i=1}^{n_1} \frac{a(1-e^{-\lambda x_i})^\delta}{(1-q)(1-a(1-q)(1-e^{-\lambda x_i})^\delta)}  \\
\frac{\partial \ell^2}{\partial \lambda \partial q} &=& \sum_{i=1}^{n_1} \frac{\eta_1 a^2 \delta x_i e^{-\lambda x_i} (1-e^{-\lambda x_i})^{2 \delta -1}}{(1-a(1-q)(1-e^{-\lambda x_i})^\delta)^2} + \sum_{j=1}^{n_2} \frac{\eta_2 a^2 \delta y_j e^{-\lambda y_j} (1-e^{-\lambda y_j})^{2 \delta -1}}{(1-a(1-q)(1-e^{-\lambda y_j})^\delta)^2}
\end{eqnarray*}

\end{document}